\let\origvec\vec
\documentclass[runningheads]{llncs}

\let\vec\origvec
\usepackage{amsmath,amssymb,amsfonts}
\usepackage{graphicx}
\usepackage{subcaption}
\usepackage{booktabs}
\usepackage{listings}
\usepackage{multirow}
\usepackage{pgfplots}
\pgfplotsset{compat=1.3}
\usepackage{algorithmic}
\usepackage{textcomp}
\usepackage{tabularx}
\usepackage{float}
\usepackage[latin1]{inputenc}
\usepackage{tikz}
\usetikzlibrary{shapes,arrows}
\newcolumntype{L}[1]{>{\raggedright\let\newline\\\arraybackslash\hspace{0pt}}m{#1}}
\newcolumntype{C}[1]{>{\centering\let\newline\\\arraybackslash\hspace{0pt}}m{#1}}
\newcolumntype{R}[1]{>{\raggedleft\let\newline\\\arraybackslash\hspace{0pt}}m{#1}}
\newcommand{\kl}{Kaby Lake}
\renewcommand{\sl}{Skylake}

\begin{document}
\title{Towards an Achievable Performance\\for the Loop Nests}
\author{
Aniket Shivam\inst{1}
\and
Neftali Watkinson\inst{1}
\and
Alexandru Nicolau\inst{1}
\and\\
David Padua\inst{2}
\and
Alexander V. Veidenbaum\inst{1}
}
\institute{
Department of Computer Science,
University of California, Irvine\\
\email{\{aniketsh,watkinso,nicolau,alexv\}@ics.uci.edu}
\and
Department of Computer Science,
University of Illinois at Urbana-Champaign\\
\email{padua@illinois.edu}
}

\authorrunning{A. Shivam et al.}
\maketitle
\begin{abstract}
Numerous code optimization techniques, including loop nest optimizations, have been developed over the last four decades.
Loop optimization techniques transform loop nests to improve the performance of the code on a target architecture, including exposing parallelism. 
%Typically, the order of optimizations/transformations is fixed because 
Finding and evaluating an optimal, semantic-preserving sequence of transformations is a complex problem.
% (phase ordering+). 
The sequence is guided using heuristics and/or analytical models and there is no way of knowing how close it gets to optimal performance or if there is any headroom for improvement.
%Hence, the performance of the generated code for the loop nests, when optimized by different code optimizers for either serial and/or parallel execution, may vary significantly.

\hspace{10pt} 
%An interesting aspect of the loop nests is their inherent behavior on a specific architecture.
This paper makes two contributions. First, it uses a comparative analysis of loop optimizations/transformations across multiple compilers to determine how much headroom may exist for each compiler. 
And second, it presents an approach to characterize the loop nests based on their hardware performance counter values and a Machine Learning approach that predicts which compiler will generate the fastest code for a loop nest. 
The prediction is made for both auto-vectorized, serial compilation and for auto-parallelization.
The results show that the headroom for state-of-the-art compilers ranges from 1.10x to 1.42x for the serial code and from 1.30x to 1.71x for the auto-parallelized code. These results are based on the Machine Learning predictions.
% In the latter case, to address the question of scalability for such loops, the model then predicts the number of threads that would be beneficial for the underlying multi-core architecture.
%We evaluate the ability of hardware performance counter to impart enough knowledge to successfully train machine learning models on two different architectures.
\end{abstract}

\section{Introduction} \label{sec:intro}
Modern architectures %present a diverse set of challenges for the compilers.
%The architectures 
have been evolving towards greater number of %processing elements (PEs) or 
cores on the chip, as well as, improving the processing capabilities of individual cores.
Each core in the current multi-core architectures includes the capability to process Single Instruction Multiple Data (SIMD) or Vector instructions.
State-of-the-art compilers, or code optimizers, use advanced loop transformation techniques to modify the loop nests so as to take advantage of these SIMD instructions.
The underlying code optimization techniques in the compilers to \textit{auto-vectorize} the loop nests\cite{Padua86,Allen87,Wolfe95} require careful analysis of data dependences, memory access patterns, etc.
%Therefore, it is expected that different compilers would transform a loop nest differently such that its performance varies on the same architecture.
Similarly, a serial version of the loop nest may be parallelized i.e. transformed such that loop iterations can be reordered and scheduled for parallel execution across the multiple cores.
These transformations are characterized as \textit{auto-parallelization} techniques\cite{Padua80,Li92,Lim98,Lim99,Lim01,Bondhugula08b,Darte12} and the end product is a multi-threaded code.

Some key transformations for optimizing loop nests\cite{Wolfe95,Kennedy01} are Distribution, Fusion, Interchange, Skewing, Tiling and Unrolling.
The best set of transformations for a given loop nest can be any possible sequence of these transformations with even repeating transformations.
Even though the compilers may have the ability to perform important loop transformations, the built-in heuristics and analytical models that drive these optimizations to determine the order and the profitability of these transformations may lead to sub-optimal results.
Evaluation studies\cite{Tournavitis09,Maleki11,Gong18} have shown that state-of-the-art compilers may miss out on opportunities to optimize the code for modern architectures.
%, including auto-vectorization and auto-parallelization of loop nests.
But a major challenge in developing heuristics and profitability models is predicting the behavior of a multi-core processor which has complex pipelines, multiple functional units, memory hierarchy, hardware data prefetching, etc.
Parallelization of loop nests involve further challenges for the compilers, since communication costs based on the temporal and spatial data locality among iterations have an impact on the overall performance too.
% Both auto-vectorization and auto-parallelization techniques evaluate loop nests for \textit{profitability} of generating SIMD code and multi-threaded code respectively.
These heuristics and models differ between compilers which leads to different quality of the generated code for the loop nests and therefore, the performance may vary significantly.
There are various compilers and domain specific loop optimizers that perform auto-vectorization and, in some cases, auto-parallelization such Intel ICC, GNU GCC, LLVM Clang, etc.
%In addition to these, there are Polyhedral Model based loop optimizers such as Polly\cite{Grosser12} that can auto-vectorize and auto-parallelize loop nests. 
By observing their relative performance one can identify relative headroom.

Embedding Machine Learning models in compilers is continuously being explored by the research community\cite{Cavazos07,Tournavitis09,Wang09,Fursin11,Stock12,Cammarota13,Watkinson17,Ashouri17}.
%\note{Add OptiScope\cite{Moseley09} and OpenTuner\cite{Ansel14}}
Most of the previous work used Machine Learning in the domain of auto-vectorization, phase-ordering and parallelism runtime settings. 
This work applies Machine Learning on a coarser level, in order to predict the most suited code optimizer - for serial as well as parallel code. %instead of predicting a component from the compiler tool-chain.
%In addition, it predicts if the auto-parallelized version of a loop nest would produce performance benefits compared to the corresponding serial version.

Previous studies have shown that hardware performance counters can successfully capture the characteristic behavior of the loop nests.
In those studies, Machine Learning models either use a mix of static features (collected from source code at compile time) and dynamic features (collected from profiling)\cite{Tournavitis09,Wang09}, or exclusively use dynamic features\cite{Cavazos07,Watkinson17,Ashouri17}. 
This work belongs to the second class and exclusively uses hardware performance counters collected from profiling a serial (-O1) version of a loop nest and uses these dynamic features as the input for the Machine Learning classifiers.
It also shows that it is feasible to use hardware performance counters from an architecture to make predictions for similar multi-core architectures.

The focus of this work is to consider state-of-the art code optimizers and then use Machine Learning algorithms to make predictions for better, yet clearly achievable performance for the loop nests using these code optimizers. This is what defines a possible headroom. We believe that recognizing the inherent behavior of loop nests using hardware performance counters and Machine Learning algorithms will present an automated mechanism for compiler writers to identify where to focus on making improvements in order to achieve better performance.
%Our approach of using Machine Learning to characterize loop nests presents an automated mechanism to assist compilers in choosing a better sequence of loop transformations.
% However, that analysis is beyond the scope of this work.
% %since we do not have access to the internals of all the compilers, it is beyond the scope of this work.

% %\note{Contribution} Predicting auto-parallelizing compiler from serial code profile. Predicting Scalabilty too.

% The rest of the paper is organized as follows.
% %In Section \ref{sec:approach}, we discuss our approach to Machine Learning based predictions of the most suited code optimizer for the loop nests.
% Section \ref{sec:expr} describes our experimental methodology and in Section \ref{sec:result} we analyze the experimental results.
% %Sections \ref{sec:priorart} discusses related work.
% We conclude the paper with Section \ref{sec:conclusion}.

%\input{approach}
\section{Experimental Methodology} \label{sec:expr}
This section describes the candidate code optimizers and the architectures that we considered for this work and methodology for conducting the experiments.

\subsection{Code Optimizers}
% There is a reasonably good choice of code optimizers that is available for most widely used architectures.
% For example, for Intel architectures, there are commercial compilers from Intel, open-source compilers like GNU GCC and Clang (LLVM).
% In addition to these, advances in Polyhedral Model techniques have showed promise in generating efficient auto-vectorized and auto-parallelized code for certain applications.
In this work we considered 4 candidate code optimizers, as shown in Table \ref{table:compilerflags}, including Polly\cite{Grosser12,Polly}, a Polyhedral Model based optimizer for LLVM. 
2 out of those 4 optimizers can perform auto-parallelization of the loop nests.
The hardware performance counters are collected using an executable generated by \texttt{icc} with flags \texttt{-O1 -no-vec}, in order to disable all loop transformations, and disable vector code and parallel code generation.
\begin{table}[b]
\centering
\begin{tabular}{|C{0.20\textwidth}|C{0.11\textwidth}|C{0.45\textwidth}|C{0.20\textwidth}|}
\hline
 \textbf{Code Optimizer} & \textbf{Version} & \textbf{Flags (Auto-Parallelization flags)} & \textbf{Auto-Parallelization} \\ \hline
 clang (LLVM)                   & 6.0.0   & -Ofast -march=native                    & No  \\ \hline
 gcc (GNU)                      & 5.4.0   & -Ofast -march=native                    & No  \\ \hline
 icc (Intel)                    & 18.0.0  & -Ofast -xHost (-parallel)               & Yes \\ \hline
 %pgcc (PGI)                     & 18.4    & -fast -tp=knl/haswell -Mllvm (-Mconcur) & Yes \\ \hline
 %PLuTo (source-to-source) + icc & 0.11.4  & \texttt{--}tile (\texttt{--}parallel)   & Yes \\ \hline
 polly                          & 6.0.0   & -O3 -march=native -polly
                                            -polly-vectorizer=stripmine  
 											-polly-tiling
                                            (-polly-parallel)                       & Yes \\ \hline
\end{tabular}
\caption{Candidate Code Optimizer and their flags}
\label{table:compilerflags}
\end{table}
\subsection{Benchmarks}
%The benchmark suites that we target in this work are specialized to test auto-vectorization and auto-parallelization capabilities of the code optimizers. 
The first benchmark suite that we use for our experiment is Test Suite for Vectorizing Compilers (TSVC) %developed by Callahan et al.\cite{Callahan88} and later translated to C 
as used by Callahan et al.\cite{Callahan88} and Maleki et al.\cite{Maleki11} for their works.
This benchmark was developed to assess the auto-vectorization capabilities of compilers.
Therefore, we only use those loop nests in the serial code related experiments.
The second benchmark suite that we collect loop nests from is Polybench\cite{Polybench}.
This suite consists of 30 benchmarks that perform numerical computations used in various domains such as linear algebra computations, image processing, physics simulation, etc.
We use Polybench for experiments involving both serial and auto-parallelized code.
We use the two largest datasets from Polybench to create our ML dataset.
In our experience, the variance of both the hardware performance counter values and the most suited code optimizer for the loop nests across the two datasets, was enough to treat them as two different loop nests.
This variance %across parameters (hardware counter values and the most suited optimizer) for the same loop nest across datasets 
can be attributed to two main reasons.
First, a different set of optimizations being performed by the optimizers based on the built-in analytical models/heuristics that drive those optimizations, since properties like loop trip counts usually vary across datasets.  
Second, the performance across datasets on an architecture with a memory hierarchy, where the behavior of memory may change on one or more levels.
This analysis was required to prevent the ML algorithms from \textit{overfitting}.
% \begin{table}[t]
% \centering
% \begin{tabular}{|c|c|c|c|C{0.15\textwidth}|C{0.18\textwidth}|}
% \hline
% Intel Code Name & Model & Frequency & Cores & Cache & Vector Extensions \\ \hline
% \kl{} & Core i7-7700K  & 4.20GHz &     4    & 32KB  L1 
% 								   			  256KB L2 
% 								   			  8MB   L3 & SSE\hspace{25pt}
% 								   			  			 AVX\hspace{25pt}
% 								   			  			 AVX2 \\ \hline
% \sl{} & Xeon Gold 6142 & 2.6 GHz & 	2 x 16  & 32KB  L1 
% 								   			  1MB   L2
% 								   			  22MB  L3 & SSE\hspace{25pt}    
% 								   			  			 AVX\hspace{25pt}
% 								   			  			 AVX2
% 								   			  			 AVX-512CD
% 								   			  			 AVX-512F \\ \hline
% \end{tabular}
% \caption{Specifications for Experimental Platforms}
% \label{table:platforms}
% \end{table}

\subsection{Experimental Platforms}
For the experiments, we used two recent Intel architectures.
The first architecture is a four-core Intel Kaby Lake Core i7-7700K.
% @ 4.20GHz with 32KB L1 cache, 256KB L2 cache and 8MB L3 cache.
This architecture supports Intel's SSE, AVX and AVX2 SIMD instruction set extensions.
The second architecture we use is a two sixteen-core Intel Skylake Xeon Gold 6142. % @ 2.6 GHz.
%Each Xeon processor has 32KB L1 cache, 1MB L2 cache, 22MB L3 cache.
The \sl{} architecture supports two more SIMD instruction set extensions, i.e., AVX-512CD and AVX-512F than the \kl{} architecture.
For the auto-parallelization related experiments, only one thread is mapped per core.

%Once the hardware performance counters are collected for the loop nests, 
We skip dynamic instruction count as a feature and normalize the rest of the hardware performance counters in terms of \textit{per kilo instructions} (PKI).
We exclude loop nests that have low value for crucial hardware performance counters such as instructions retired.
From our experiments, we discovered two interesting correlations among hardware performance counters and the characteristic behavior of the loop nests. 
%We use these hardware performance counters values as features to make prediction for both \kl{} and \sl{} machines.
% In order to collect the execution times, we run each loop from each candidate at least 3 times and monitor variance.
First, the hardware performance counters values from \kl{} architecture (after disabling loop transformations and vector code generation) were sufficient to get well trained ML model to make predictions for a similar architecture like the \sl{} architecture.
Second, for predicting the most suited candidate for serial code and for the auto-parallelized code for a loop nest, the same set of hardware performance counters, collected from profiling a serial version, can be used to train the ML model and achieve satisfactory results.

%For \kl{} processor, experiments are performed with a total of 4 threads, whereas for \sl{} processor it is 32 threads.
% In total this architecture has 64 hardware threads, i.e., 2 hardware threads per core. 
% Therefore, for the experiments regarding the scalability of the loop nests, we analyze performance for 1 thread, 32 threads and 64 threads.
% 64 threads essentially means exposing maximum amount of parallelism available, whereas 32 threads essentially means 1 thread per core, i.e., half of the available parallelism.

\subsection{Machine Learning Model Evaluation}
For training and evaluating our Machine Learning model, we use Orange\cite{Demvsar13}.
We use Random Forest (RF) as the classifier for all the experiments.
% We evaluated the results from Support Vector Machine (SVM) models and found that these models performed either similar or worse than Random Forest across experiments, therefore we have excluded SVM from the analysis.   
We randomly partition our dataset into Training dataset (75\%) and Validation dataset (25\%).
%The training dataset makes up 75\% of the entire dataset.
The training dataset allows us to train and tune the ML models.
We evaluate our trained models on Accuracy and Area Under Curve (AUC).
Whereas, the validation dataset is a set of unseen loop nests that we use to make predictions.
For serial code experiments% that include loop nests from TSVC and Polybench benchmark suites
, there are 209 instances (loop nests) in the training dataset and 69 instances in the validation dataset.
For auto-parallelized code experiments% which only include loop nests from Polybench benchmark suite
, there are 147 instances in the training dataset and 49 instances in the validation dataset.
%For evaluating the performance of the ML predictions, we refer to the database that stores the execution time for the loop nests when optimized by different code optimizers.
The predicted optimizer's execution time as compared to that of the most suited optimizer's execution time will be same in case of correct predictions and higher in case of mispredictions.

We repeat our ML experiments thrice in order to validate our results, i.e., we randomly split the dataset, train new ML models and then make the predictions.
%Since the dataset is randomly split in each iteration, there might be repetitions of loop nests/instances.
We take into account the unique instances from the three validation datasets for measurements. 
% The results are presented for all three validation sets combined, minus the repetitions. Thus the total number of loops is approximately three times larger than size of the validation set mentioned above.
Therefore, the number of instances differ between similar experiments.

%As mentioned before in Section \ref{section:ML_setup}, 

% \begin{figure*}[ht]
%   \centering
%   \includegraphics[width=0.75\paperwidth]{figs/mC_framework.png}
%   \caption{mCompiler Framework} 
%   \label{fig:framework} 
% \end{figure*}

\section{Experimental Analysis} \label{sec:result}
%This section presents the experimental results and discusses the findings from the ML models.
For evaluating the results, we calculate the speedup of ML predictions over candidate code optimizers, i.e., the speedup obtained if the code optimizer recommended by the ML model was used to optimize loop nests instead of a candidate.%\note{Add note about distributions.}
%Serial-KabyLake
\begin{figure}[ht]
\captionsetup{justification=centering}
\begin{subfigure}{0.49\textwidth}
\begin{tikzpicture}
	\begin{axis}[
		x tick label style={font=\scriptsize,text width=1cm,align=center},
	    width = 0.98\textwidth, height = 4.5cm,
		ybar=0pt,
    	/pgf/bar shift=0pt,
	    ymajorgrids = true,
		ylabel=GeoMean Speedups of\\ML predictions,
	    xtick={0,1,2,3},
	    xticklabels={Over Clang, Over GCC, Over ICC, Over Polly},         
	    %x tick label style={rotate=45, anchor=east, align=left},
        nodes near coords,
		%nodes near coords style={text=black},
		/pgf/bar shift=0pt,
	    ymax=1.6,
	    ymin=0.75,
	    ytick={0,0.20,...,2},
	    enlarge x limits={abs=0.825},
	    nodes near coords,
	    %nodes near coords style={text=black},
	    ylabel style={font=\scriptsize,align=center},
	  ]	           
		\addplot[red,fill] coordinates {(0, 1.3910862577)};
		\addplot[green,fill] coordinates {(1, 1.4156938278)};
		\addplot[blue!65,fill] coordinates {(2, 1.1852788349)};
		\addplot[black!45,fill] coordinates {(3, 1.2572846128)};
		\draw [dashed] (-1,1) -- (4,1);
	\end{axis}
\end{tikzpicture}        
\caption{Predictions against individual compilers on \kl{}}
\label{graph:serialpred_kl}
\end{subfigure}
\hfill
\begin{subfigure}{0.49\textwidth}
\begin{tikzpicture}
	\begin{axis}[
		x tick label style={font=\scriptsize,text width=1cm,align=center},
	    width = \textwidth, height = 4.5cm,
		ybar=0pt,
    	/pgf/bar shift=0pt,
	    ymajorgrids = true,
		ylabel=GeoMean Speedups of\\ML predictions,
	    xtick={0,1,2,3},
	    xticklabels={Over Clang, Over GCC, Over ICC, Over Polly},         
	    %x tick label style={rotate=45, anchor=east, align=left},
        nodes near coords,
		%nodes near coords style={text=black},
		/pgf/bar shift=0pt,
	    ymax=1.6,
	    ymin=0.75,
	    ytick={0,0.20,...,2},
	    enlarge x limits={abs=0.825},
	    nodes near coords,
	    %nodes near coords style={text=black},
	    ylabel style={font=\scriptsize,align=center},
	  ]
		\addplot[red,fill] coordinates {(0, 1.3430505364)};
		\addplot[green,fill] coordinates {(1, 1.3685332892)};
		\addplot[blue!65,fill] coordinates {(2, 1.1039997676)};
		\addplot[black!45,fill] coordinates {(3, 1.3501688596)};
		\draw [dashed] (-1,1) -- (4,1);
	\end{axis}
\end{tikzpicture}        
\caption{Predictions against individual compilers on \sl{}}
\label{graph:serialpred_sl}
\end{subfigure}
\begin{subfigure}{0.49\textwidth}
\begin{tabular}{ccccccc}
                        & \multicolumn{1}{c}{} & \multicolumn{4}{c}{\textbf{Predicted}}                                                                                         &  \\
						\multirow{6}{*}{\textbf{Actual}} &                        &   Clang                 &         GCC              &         ICC              &          Polly             &               \\ \cline{3-6}
                        
                        & \multicolumn{1}{c|}{Clang} & \multicolumn{1}{c|}{\textbf{5}} & \multicolumn{1}{c|}{0} & \multicolumn{1}{c|}{13} & \multicolumn{1}{c|}{7} &  25 \\ \cline{3-6}
                        
                        & \multicolumn{1}{c|}{GCC} & \multicolumn{1}{c|}{1} & \multicolumn{1}{c|}{\textbf{0}} & \multicolumn{1}{c|}{13} & \multicolumn{1}{c|}{2} &  16 \\ \cline{3-6}
                        
                        & \multicolumn{1}{c|}{ICC} & \multicolumn{1}{c|}{2} & \multicolumn{1}{c|}{0} & \multicolumn{1}{c|}{\textbf{96}} & \multicolumn{1}{c|}{2} & 100 \\ \cline{3-6}
                        
                        & \multicolumn{1}{c|}{Polly} & \multicolumn{1}{c|}{2} & \multicolumn{1}{c|}{0} & \multicolumn{1}{c|}{15} & \multicolumn{1}{c|}{\textbf{14}} & 31 \\ \cline{3-6}
                        
                        &      &  10                &        0               &         137              &        25              &         172             
%Acc=67%, Maj=58%, AUC=71
\end{tabular}
\caption{Confusion Matrix for \kl{}}
\label{table:confusionserial_kl}
\end{subfigure}
\hfill
\begin{subfigure}{0.49\textwidth}
\begin{tabular}{ccccccc}
                        & \multicolumn{1}{c}{} & \multicolumn{4}{c}{\textbf{Predicted}}                                                                                         &  \\
						\multirow{6}{*}{\textbf{Actual}} &                        &   Clang                 &         GCC              &         ICC              &          Polly             &               \\ \cline{3-6}
                        
                        & \multicolumn{1}{c|}{Clang} & \multicolumn{1}{c|}{\textbf{4}} & \multicolumn{1}{c|}{4} & \multicolumn{1}{c|}{14} & \multicolumn{1}{c|}{1} &  23 \\ \cline{3-6}
                        
                        & \multicolumn{1}{c|}{GCC} & \multicolumn{1}{c|}{4} & \multicolumn{1}{c|}{\textbf{10}} & \multicolumn{1}{c|}{11} & \multicolumn{1}{c|}{3} &  28 \\ \cline{3-6}
                        
                        & \multicolumn{1}{c|}{ICC} & \multicolumn{1}{c|}{4} & \multicolumn{1}{c|}{5} & \multicolumn{1}{c|}{\textbf{68}} & \multicolumn{1}{c|}{4} & 81 \\ \cline{3-6}
                        
                        & \multicolumn{1}{c|}{Polly} & \multicolumn{1}{c|}{0} & \multicolumn{1}{c|}{5} & \multicolumn{1}{c|}{8} & \multicolumn{1}{c|}{\textbf{15}} & 28 \\ \cline{3-6}
                        
                        &      &  12                &        24               &         101              &        23              &         160           
%Acc=61%, Maj=50%, AUC=68
\end{tabular}
\caption{Confusion Matrix for \sl{}}
\label{table:confusionserial_sl}
\end{subfigure}
\caption{Speedup of Predictions for Serial Code}
\end{figure}
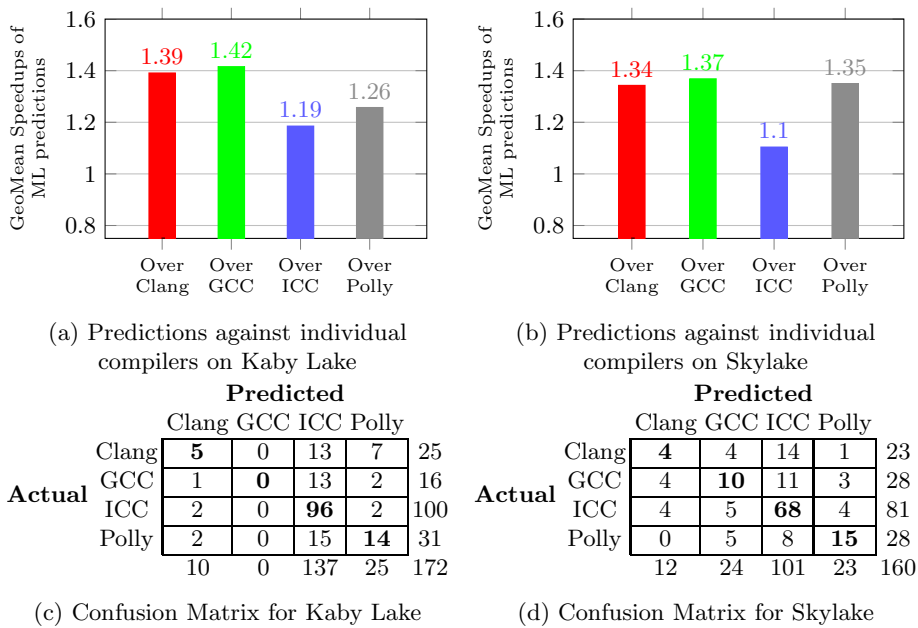

\subsection{Predicting the Most Suited Code Optimizer for Serial Code}
Fig. \ref{graph:serialpred_kl} and Fig. \ref{graph:serialpred_sl} show the results for the performance gains from the predictions for the \kl{} and \sl{} architectures, respectively.  These predicted gains can be viewed as the achievable headroom for each compiler.
On the validation dataset, RF classifier predicted with an overall accuracy of 67\% for \kl{} and 61\% for \sl{} as shown in the confusion matrices in Fig. \ref{table:confusionserial_kl} and Fig. \ref{table:confusionserial_sl} respectively.

Across both architectures, Intel compiler performs well on majority of the loop nests.
Therefore, the Majority Classifier predicted ICC with 58\% overall accuracy for \kl{} and 50\% overall accuracy for \sl{}.
% Based on the ML predictions, the geometric mean of the performance gains are up to 1.39x over Clang, 1.42x over GCC, 1.19x over ICC and 1.35x over Polly across architectures.
The distribution of performance of the ML predictions compared to ICC, the maximum performance gain on a loop nest was 27x, whereas the maximum slowdown was 0.2x.

% We ranked the features using the same RF classifier that is used for making predictions.
% The top features included hardware performance counters that are related to stall cycles induced by memory accesses, L3 cache misses, Cycles Per Instruction (CPI), D-TLB load misses, L1-D cache misses and Floating Point(FP) arithmetic instruction count.
%
%:%s/^/\=printf('(%-4d,', line('.'))
\subsection{Predicting the Most Suited Code Optimizer for Auto-Parallelized Code}
%Parallel-KabyLake
\begin{figure}[t]
\captionsetup{justification=centering}
\begin{subfigure}{0.29\textwidth}
\begin{tikzpicture}
	\begin{axis}[
		x tick label style={font=\scriptsize,text width=1cm,align=center},
	    width = 0.98\textwidth, height = 5cm,
		ybar=0pt,
    	/pgf/bar shift=0pt,
	    ymajorgrids = true,
		ylabel=GeoMean Speedups of\\ML predictions,
	    xtick={0,1},
	    xticklabels={{Over ICC}, {Over Polly}},         
	    %x tick label style={rotate=45, anchor=east, align=left},
        nodes near coords,
		%nodes near coords style={text=black},
		/pgf/bar shift=0pt,
	    ymax=1.8,
	    ymin=0.9,
	    ytick={0,0.20,...,2},
	    enlarge x limits={abs=0.825},
	    nodes near coords,
	    %nodes near coords style={text=black},
	    ylabel style={font=\scriptsize,align=center},
	  ]				           
		\addplot[blue!65,fill] coordinates {(0, 1.4648890881) };
		\addplot[black!45,fill] coordinates {(1, 1.5073725847)};
		\draw [dashed] (-1,1) -- (2,1);
	\end{axis}
\end{tikzpicture}        
\caption{Predictions against individual compilers\\on \kl{}}
\label{graph:parpred_kl}
\end{subfigure}
\begin{subfigure}{0.29\textwidth}
\begin{tikzpicture}
	\begin{axis}[
		x tick label style={font=\scriptsize,text width=1cm,align=center},
	    width = 0.99\textwidth, height = 5cm,
		ybar=0pt,
    	/pgf/bar shift=0pt,
	    ymajorgrids = true,
		ylabel=GeoMean Speedups of\\ML predictions,
	    xtick={0,1},
	    xticklabels={{Over ICC}, {Over Polly}},         
	    %x tick label style={rotate=45, anchor=east, align=left},
        nodes near coords,
		%nodes near coords style={text=black},
		/pgf/bar shift=0pt,
	    ymax=1.8,
	    ymin=0.9,
	    ytick={0,0.20,...,2},
	    enlarge x limits={abs=0.825},
	    nodes near coords,
	    %nodes near coords style={text=black},
	    ylabel style={font=\scriptsize,align=center},
	  ]				           
		\addplot[blue!65,fill] coordinates {(0, 1.7122834354) };
		\addplot[black!45,fill] coordinates {(1, 1.304809132)};
		\draw [dashed] (-1,1) -- (2,1);
	\end{axis}
\end{tikzpicture}        
\caption{Predictions against individual compilers\\on \sl{}}
\label{graph:parpred_sl}
\end{subfigure}
\begin{subfigure}{0.40\textwidth}
\begin{tabular}{ccccc}
                                 & \multicolumn{1}{c}{}       & \multicolumn{2}{l}{\textbf{Predicted}}                              &     \\
                                 &                            & ICC                              & Polly                            &     \\ \cline{3-4}
									\multirow{2}{*}{\textbf{Actual}} & \multicolumn{1}{c|}{ICC}   & \multicolumn{1}{c|}{\textbf{65}} & \multicolumn{1}{c|}{9}          &  74 \\ \cline{3-4}
                                 & \multicolumn{1}{c|}{Polly} & \multicolumn{1}{c|}{8}          & \multicolumn{1}{c|}{\textbf{33}} &  41 \\ \cline{3-4}
                                 &                            &            73                    &    42                            & 115 \\
                                 &&&&
%Acc=85%, Maj=64%, AUC=92
\end{tabular}
\begin{tabular}{ccccc}
                                 & \multicolumn{1}{c}{}       & \multicolumn{2}{l}{\textbf{Predicted}}                              &     \\
                                 &                            & ICC                              & Polly                            &     \\ \cline{3-4}
									\multirow{2}{*}{\textbf{Actual}} & \multicolumn{1}{c|}{ICC}   & \multicolumn{1}{c|}{\textbf{40}} & \multicolumn{1}{c|}{14}          & 54  \\ \cline{3-4}
                                 & \multicolumn{1}{c|}{Polly} & \multicolumn{1}{c|}{16}          & \multicolumn{1}{c|}{\textbf{38}} & 54  \\ \cline{3-4}
                                 &                            & 56                               & 52                               & 108
%Acc=72%, Maj=50%, AUC=76
\end{tabular}
\caption{Confusion Matrix for\\\kl{} (top) and\\\sl{} (bottom)}
\label{table:confusionpar}
\end{subfigure}
\caption{Speedup of Predictions for Auto-Parallelized Code}
\end{figure}
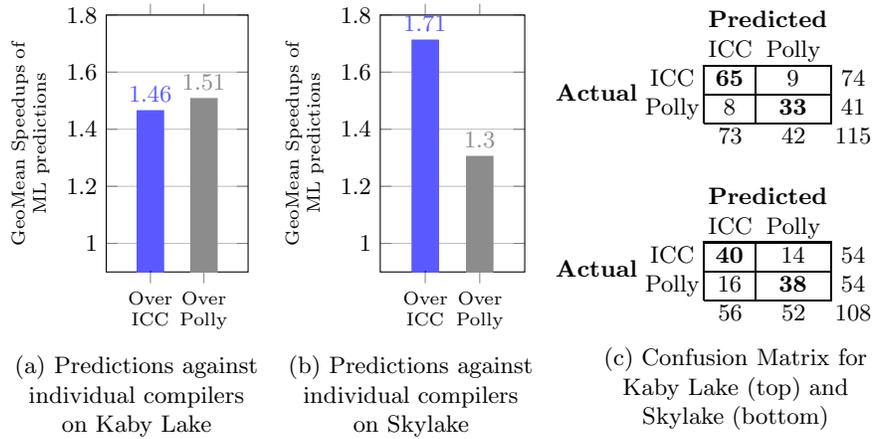
For the auto-parallelization experiments, there are only two candidates: ICC and Polly.
The RF classifier predicted with an overall accuracy of 85\% for \kl{} and 72\% for \sl{} as shown in Fig. \ref{table:confusionpar}.
Since the validation dataset was well balanced for the two targets, the Majority Classifier produced an overall accuracy of 64\% for \kl{} and 50\% for \sl{}.
% The performance gains (as shown in Fig. \ref{graph:parpred_kl} for \kl{} and Fig. \ref{graph:parpred_sl} for \sl{}) from the ML predictions are evidence of the greater overall accuracy of the models.
% The geometric mean of the performance gains from the ML predictions over ICC was up to 1.71x and up to 1.51x over Polly across architectures.
Based on the distribution of performance of the ML predictions, when compared to ICC, the maximum gain on a loop nest was 91x whereas the maximum slowdown was 0.09x.

\section{Overall Analysis and Discussion}
The performance gain from the ML predictions over the candidate code optimizers range from 1.10x to 1.42x for the serial code and from 1.30x to 1.71x for the auto-parallelized code across two multi-core architectures.
%The performance gain would be even larger if it was not for the mispredictions from the ML models.
%Based on all the experimental results, for the both serial and auto-parallelized code predictions across architectures, we found that certain hardware performance counters were prominent for making such predictions.
Counters related to Cycles Per Instruction (CPI), D-TLB, memory instructions, cache performance (L1, L2 and L3) and stall cycles were crucial indicators of the inherent behavior of the loop nests.
%Another interesting observation from the experiments was the variance in performance between the code optimizers for the loop nests.

On analyzing the validation datasets for serial code experiments, we found that on an average for 95\% of the loop nests, there was at least 5\% performance difference between the most suited code optimizer and the worse suited code optimizer. 
For auto-parallelized code experiments, on an average for 91.5\% of the loop nests, there was at least 5\% performance difference between the most suited code optimizer and the worse suited code optimizer.

On the other hand, for the serial code experiments, for 68\% of the loop nests, there was at least 5\% performance difference between the most suited code optimizer and the second most suited code optimizer.
That suggests that for the remaining 32\% of the loop nests, it would be harder to make a distinction between the most suited code optimizer and the second one.
Since the ML models' overall accuracy are 67\% for \kl{} and 61\% for \sl{}, we can infer that they are doing very well on the loop nests that have a clear distinction about the most suited code optimizer.
\let\thefootnote\relax\footnote{\textbf{Acknowledgments.} This work was supported by NSF award XPS 1533926.}
%
%
%
%An interesting note about our models is that while our overall accuracy is over 60\%, the average accuracy (accuracy per target divided by number of targets) is 40\% for Kaby Lake and  47.6\% for Skylake. From a statistical point of view, there's a simple explanation for this difference. For Kaby Lake, ICC is the actual value for the majority of the loops, which overwhelms the model. Skylake, on the other hand, has an equal distribution between loops that ICC does best and loops that doesn't. Therefore the classifier is able to model a pattern for the other compilers even when that means sacrificing the overall accuracy. An equally distributed dataset would solve this problem, but for that we need to find more samples of loop nests where GCC, Clang and Polly do better than ICC.
%From the point of view of the compilers, we can argue that Clang, GCC and ICC are in the end similar compilers for which many of the loops that each one of them do best, it is just by a small edge (which is also why we still gain performance even when a compiler is mispredicted). There are loops where there's a clear winner among those three due to a peculiarity of their optimizing algorithm. We decided to include both types of loops (where the compiler has a slight edge and where there's a clear winner) to help the classifier find a pattern. Polly is a separate case since it applies to very specific loop patterns. The difficulty with it is finding loop nests that benefit from Polly.

%\input{related_work}
%\input{summary}

\bibliographystyle{abbrv}
\bibliography{bibliography}
\end{document}